# Layer-Resolved Many-Electron Interactions in Delafossite PdCoO$_2$ from Standing-Wave Photoemission Spectroscopy

Qiyang Lu[1,2,3,4 &,#], Henrique Martins[2,5 &], J. Matthias Kahk[6 &], Gaurab Rimal[7], Seongshik Oh[7], Inna Vishik[5], Matthew Brahlek[8], William C. Chueh[3,4], Johannes Lischner[6,9 #], Slavomir Nemsak[2 #]

[1] School of Engineering, Westlake University, Hangzhou, Zhejiang, 310024 China

[2] Advanced Light Source, Lawrence Berkeley National Laboratory, Berkeley, CA 94720, US

[3] Department of Materials Science and Engineering, Stanford University, Stanford, CA 94306, US

[4] Stanford Institute for Materials and Energy Science, SLAC National Accelerator Laboratory, Menlo Park, CA, USA

[5] Department of Physics, University of California Davis, Davis, CA, 95616 US

[6] Department of Materials, Imperial College London, South Kensington Campus, London SW7 2AZ, UK

[7] Department of Physics and Astronomy, Rutgers, The State University of New Jersey, Piscataway, New Jersey 08854, US

[8] Materials Science and Technology Division, Oak Ridge National Laboratory, Oak Ridge, TN 37830, US

[9] Thomas Young Center for Theory and Simulation of Materials

[&] These authors contributed equally

[#] Corresponding author: luqiyang@westlake.edu.cn, j.lischner@imperial.ac.uk & snemsak@lbl.gov

**Abstract**

When a three-dimensional material is constructed by stacking different two-dimensional layers into an ordered structure, new and unique physical properties can emerge. An example is the delafossite PdCoO$_2$, which consists of alternating layers of metallic Pd and Mott-insulating CoO$_2$ sheets. To understand the nature of the electronic coupling between the layers that gives rise to the unique properties of PdCoO$_2$, we revealed its layer-resolved electronic structure combining standing-wave X-ray photoemission spectroscopy and *ab initio* many-body calculations. Experimentally, we have decomposed the measured valence band spectrum into contributions from Pd and CoO$_2$ layers. Computationally, we find that many-body interactions in Pd and CoO$_2$ layers are highly different. Holes in the CoO$_2$ layer interact strongly with charge-transfer excitons in the same layer, whereas holes in the Pd layer couple to plasmons in the Pd layer. Interestingly, we find that holes in states hybridized across both layers couple to both types of excitations (charge-transfer excitons or plasmons), with the intensity of photoemission satellites being proportional to the projection of the state onto a given layer. This establishes satellites as a sensitive probe for inter-layer hybridization. These findings pave the way towards a better understanding of complex many-electron interactions in layered quantum materials.

**Introduction**

Materials can be categorized into two broad classes depending on the strength of correlations among the electrons. Materials with weak or modest correlations can be successfully described by band structure theory based on the assumption of delocalized electrons that interact through long-ranged Coulomb interactions. In contrast, electrons in strongly correlated materials can be localized in space and are coupled through strong short-ranged Hubbard-type interactions. While most materials fall into one or the other category, it is intriguing to consider what will happen when the two regimes coexist side-by-side in the same crystal structure.

Delafossite oxides, such as PdCoO$_2$ or PtCoO$_2$[1,2], have recently emerged as a versatile platform for studying exotic quantum phenomena[3]. These materials can be viewed as heterostructures composed of metallic layers consisting of triangularly coordinated metal atoms (such as Pd or Pt) separated by Mott insulating layers of edge-sharing CoO$_6$ octahedra[4] and are therefore ideal candidates for studying the interplay of weakly and strongly correlated electrons. As a consequence of the two-dimensional confinement, electron conduction in $PdCoO_2$ or $PtCoO_2$ is highly anisotropic with in-plane conductivities in the metallic sheets being ~$10^3$ times higher than out-of-plane conductivities[5,6] Remarkably, the in-plane conductivity of these materials is the highest reported among all known oxides[7]. In addition, these materials exhibit numerous exotic phenomena, including extremely large magnetoresistance[8], quantum oscillations[7,9] and hydrodynamic electron flow[10]. Some studies also explored the possibility of using $PdCoO_2$ as a electrocatalyst for the oxygen evolution reaction (OER)[11] and hydrogen evolution reaction (HER)[12,13].

The first model of the electronic structure of $PdCoO_2$ was proposed by Rodgers and coworkers[14]. They suggested that Pd 5s and $4dz^2$ orbitals hybridize and that the resulting ($5s+4dz^2$) orbitals couple to the oxygen 2p orbitals to form broad bonding and anti-bonding sigma bands. In contrast, the ($5s-4dz^2$) Pd orbitals as well as the other Pd d-orbitals are non-bonding with respect to the O atoms, but are strongly oriented in the direction of the neighboring Pd atoms giving rise to the formation of metallic bands. This analysis was later corroborated by detailed first-principles density-functional theory (DFT) calculations[15,16].

Experimentally, photoemission spectroscopy has been used as powerful tool to gain insight into the electronic structure of the delafossite oxides. Two recent studies used angle-resolved photoemission spectroscopy (ARPES) to measure the electronic quasiparticle band structure (with the quasiparticle being a hole with an associated screening cloud of electron-hole pair excitations) and Fermi surfaces of $PtCoO_2$[5] and $PdCoO_2$[17] and found good agreement with first-principles calculations. However, standard photoemission techniques cannot provide direct information about the layer-specific electronic structure or shed light on the interaction of weakly correlated electrons in the Pd layers and the strongly correlated electrons in the $CoO_2$ sheets. To overcome this problem, we use the depth specificity of standing-wave X-ray photoemission spectroscopy (SW-XPS), which has been successfully used to study complex oxide quantum materials in the form of multilayers, as well as single crystalline materials.[18–23] In single crystals, an X-ray standing wave is created by using Bragg reflection off atomic planes, which results in enhanced photoemission intensity from atoms whose positions coincide with the maxima (or antinodes) of the standing wave (see Figure 1). By measuring the change in the photoemission intensities as function of the incident angle, it is possible to decompose the valence band spectrum into contributions from the individual layers as we demonstrate below.

Besides band structure contributions (direct photoemission process), the layer-resolved valence band photoemission spectra also contain fingerprints of many-body interactions (manifested through satellite peaks) resulting from the coupling of valence electron holes created by the photoemission process to other excitations, such as plasmons or excitons. To identify the direct and satellite components of the layer resolved experimental spectra, we have calculated layer-resolved photoelectron spectra of $PdCoO_2$ using first principles GW+cumulant theory[24–28], which provides an accurate description of plasmon satellites in materials. In addition, we have simulated photoemission from the Pd and $CoO_2$ layers using a finite cluster approach[29,30] which captures effects of strong electron correlations, such as multiplets. The two radically different theoretical methods are found to agree well with each other, and also agree well with experiment, which allows us to clearly identify and interpret the layer-resolved satellite features. As such, our study brings a deeper understanding of these complicated many-body interactions in layered quantum materials where weakly and strongly correlated electrons coexist.

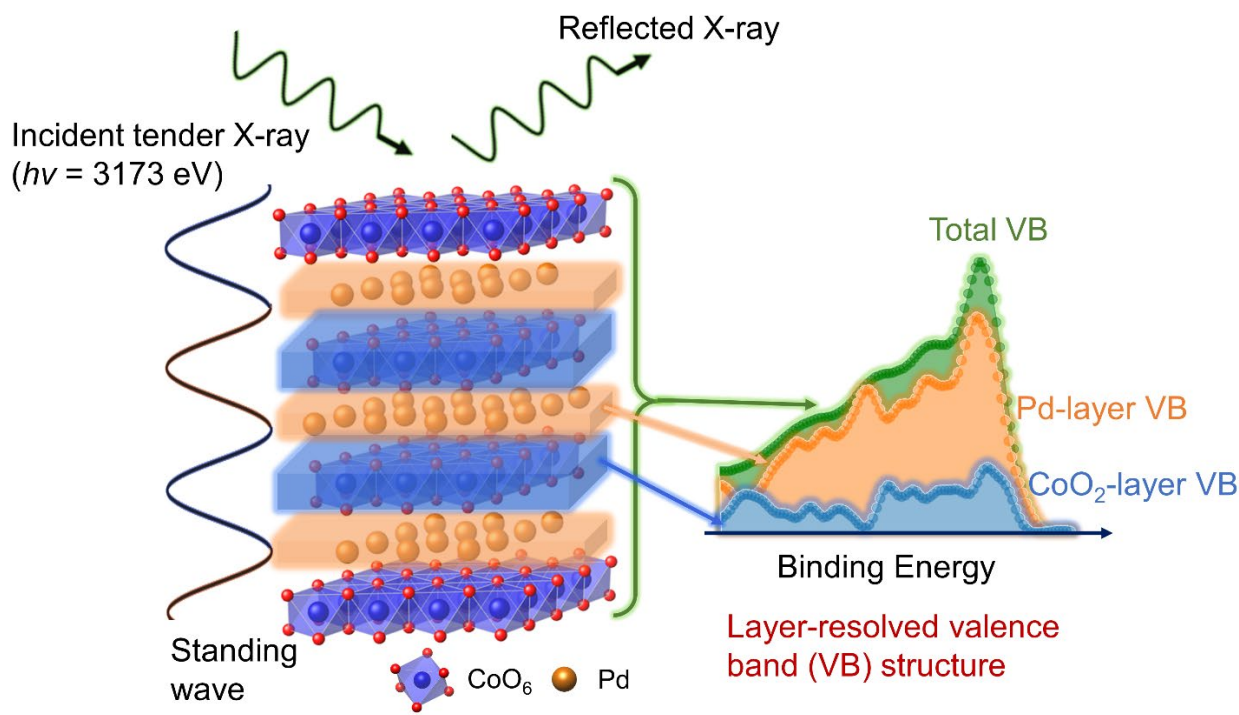

Figure 1: Schematic figure describing the measurement of the layer-resolved electronic structure of $PdCoO_2$ with standing-wave X-ray photoemission spectroscopy (SW-XPS). A standing wave is excited in the material using tender/hard X-rays with a photon energy of 3173 eV. While traditional angle-integrated photoemission spectroscopy only measures the total valence band (VB) structure (shown in green, denoted by "Total VB"), SW-XPS allows the separation of the total VB structure into contributions from the Pd layer (shown in orange, denoted by "Pd-layer VB") and from the $CoO_2$ layer (shown in blue, denoted by "$CoO_2$-layer VB").

## Results and Discussion

We first present the experimental results and computational analysis on core level spectroscopy of $PdCoO_2$. In contrast to valence band (VB) photoemission which probes delocalized electronic states, core-level wavefunctions are highly localized and the resulting core-level photoemission spectrum in layered oxides is naturally layer-specific. Core-level spectra of Co 2p and Pd 3d using hard X-rays with photon energies of 4 keV as well as spectra obtained using a lab-source soft X-ray (1486.6 eV) are shown in Figure 2. The Co 2p spectrum exhibits one satellite peak ~10 eV above the main peak and a shoulder which is ~2 eV above the main peak. The Pd 3d spectrum exhibits a small satellite peak which is separated by ~8 eV from the main line.

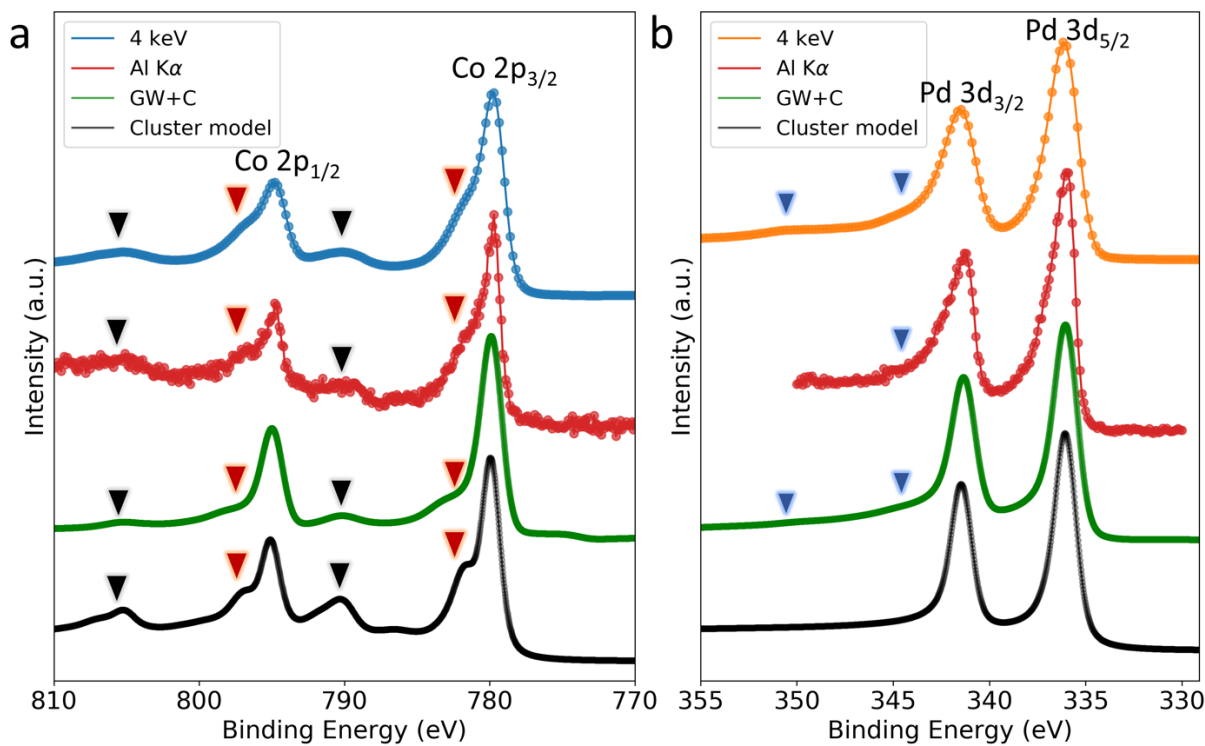

Figure 2: Core-level photoemission spectra of PdCoO$_2$. (a) Co 2p spectra collected by using both tender/hard X-rays ($hv$ = 4 keV) and monochromated Al Kα X-rays ($hv$ = 1.486 keV) are compared to cluster model (black) and GW+cumulant (green) results. The black triangles indicate charge-transfer satellites while the red triangles indicate the position of the feature resulting from non-local screening. (b) Equivalent graphs for the Pd 3d spectrum. The blue triangles indicate plasmon satellites. Note that the cluster model calculation does not reproduce the plasmon satellites while the GW+cumulant approach does.

In the Co 2p core level spectrum, the 2 eV shoulder feature has also been observed in previous works on PdCoO$_2$, and also in Li$_x$CoO$_2$. Noh *et al.* attributed this feature to the existence of surface contaminants[17]. However, we find that the shoulder has roughly the same relative intensity compared to the main peak in spectra collected using both soft and hard X-rays. As the different photon energies result in very different probing depths (the inelastic mean free path is ~1.3 nm for Al Kα excited photoelectrons and ~4.1 nm for hν=4 keV excited photoelectrons), the shoulder cannot be simply assigned to a surface contaminant, and an alternative explanation needs to be sought.

In order to gain insight into the origin of the satellite features, we have calculated the Co 2p core level spectrum in PdCoO$_2$ using a cluster model (see Methods section). In this approach, the many-electron Hamiltonian of a small cluster of atoms is diagonalized which allows the description of short-ranged strong correlation effects with high accuracy. The cluster model reproduces both the 10 eV satellite and the 2 eV shoulder features resulting in good agreement between theory and experiment, as shown in Fig. 2(a). In agreement with previous work[29], we assign the 10 eV satellite to a charge transfer excitation between Co and O atoms induced by the presence of the core hole. The 2 eV shoulder feature is only reproduced by the calculations when a cluster consisting of two connected CoO$_6$ octahedra is considered. This suggests that the shoulder feature originates from charge transfer between the two octahedra and is therefore a consequence of non-local screening[31,32].

We have also performed first principles calculations to model photoemission from PdCoO$_2$ using the GW+cumulant (GW+C) approach. In our calculations (see methods section), the electronic structure is described using a plane wave/pseudopotential approach, and therefore the Co 2p core electrons are not explicitly included. Instead, we have considered the calculated spectral function of a specific valence state that is predominantly localized on the Co atoms based on Löwdin analysis. Assuming that the spectral function of this valence state has a similar shape as that of a Co 2p electron, we have constructed a theoretical core-level spectrum by adding up two copies of the valence band spectral function with one copy shifted by the atomic spin-orbit splitting and weighted by the theoretical intensity ratio from the multiplicities of the *j*-states. This approximate Co 2p core level spectrum from GW+C theory is shown in Figure 2(a), and it accurately reproduces both the charge-transfer satellite and the shoulder feature.

In a similar manner, we have used GW+C and cluster model calculations to predict the appearance of the Pd 3d core level spectrum in PdCoO$_2$. The experimental Pd 3d core level spectrum shows weak but detectable satellites at approximately 8 eV above the main peaks of the spin-orbit doublet, as shown in Fig. 2(b). These satellites are reproduced by the GW+C calculation, but they are not present in the spectrum obtained from the cluster approach. This suggests that the 8 eV satellite arises from coupling to a collective excitation, *e.g.* a plasmon, in $PdCoO_2$.

Next, we turn to the SW-XPS results of both core-level and valence band spectroscopy. Figure 3(a) shows the measured photoemission intensities of Co 2p, O 1s and Pd 3d core levels as a function of the incident angle $\theta$, which are conventionally referred to as rocking curves (RCs). Interestingly, the RCs of core levels from atoms in the $CoO_2$ layers, *i.e.* Co 2p and O 1s, have maximum intensity near $\theta = 19.5$ degree which corresponds to the (0003) Bragg angle. In contrast, the photoemission intensity from the Pd layer shows a minimum at this angle. To understand this finding, we have calculated the depth-dependent electric field as function of $\theta$ using dynamical diffraction theory, see Figure 3(b). Near the (0003) Bragg angle, the formation of a standing wave can be observed. At $\theta = 19.5$ degree, the standing wave nodes coincide with the positions of the Pd layers explaining the weakening of the Pd 3d core level intensity, while the maxima of the electric field distribution are located at the position of the $CoO_2$ layers which gives rise to intense photoemission from the corresponding Co and O core levels. We note that the peaks of the Co 2p and O 1s RCs are somewhat broader than those of the Pd 3d RCs. This is likely a consequence of a higher concentration of defects in the $CoO_2$ layers.

Quantitatively, the rocking curves $I(\theta)$ can be described *via*

$$I(\theta) = 1 + R(\theta) + 2\sqrt{R(\theta)} f_H \cos(v(\theta) - 2\pi P_H), \qquad \text{(Eqn. 1)}$$

where $R(\theta)$ is the reflectivity of the $PdCoO_2$ thin film and, $v(\theta)$ is the phase of the standing wave. $f_H$ and $P_H$ denote the coherent fractions and coherent positions, respectively[33,34]. Figure 3(a) shows that the calculated RCs agree well with the measured ones. For the calculated RCs, we used $f_H = 1$ for both Pd and Co, and coherent positions of 0.75 and 0.25 for Pd and Co, respectively. The $f_H$ and $P_H$ values are directly obtained from the crystal structure of $PdCoO_2$. To model the oxygen RCs, we add the contributions from both oxygen atoms in the $CoO_2$ layer using $P_H$ values of 0.09 and 0.41 and coherent fractions $f_H$ of 0.5.

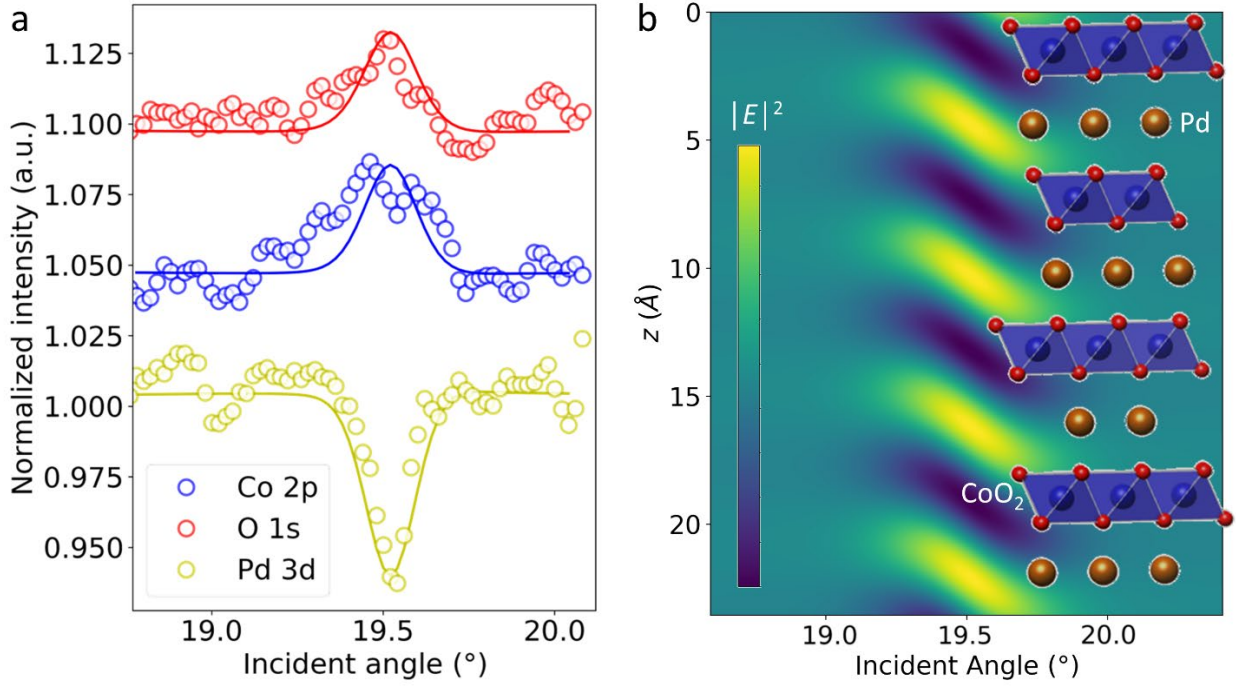

Figure 3: Experimental and simulated rocking curves of core-level spectra of PdCoO$_2$. (a) Rocking curves of Pd 3d, Co 2p and O 1s near the (0003) Bragg peak. Experimental data points are shown by circles while the lines are simulated rocking curves obtained from dynamical diffraction theory. (b) Calculated electrical field strength ($|E|^2$) as a function of depth and incident angle.

In order to gain insight into the layer-resolved valence band (VB) electronic structure of PdCoO$_2$, we measured the VB photoemission spectrum at different incident angles near the (0003) Bragg peak. To decompose the total VB spectrum into contributions from the Pd and the CoO$_2$ layers, we express the measured VB intensity $I_{VB}(E_b, \theta)$ (with $E_b$ denoting the binding energy) in terms of the layer-resolved core-level intensities $I_j(\theta)$ according to

$$I_{VB}(E_b, \theta) = \sum_j \rho_j(E_b) I_j(\theta), j = Pd \text{ or } CoO_2 \quad \text{(Eqn. 2)}$$

where $\rho_{Pd}(E_b)$ and $\rho_{CoO_2}(E_b)$ denote the layer-resolved contributions to the valence band spectrum obtained from a least-square fit. $I_{Pd}(\theta)$ and $I_{CoO_2}(\theta)$ denote the rocking curve intensity of core level spectrum from each layer. Note that either the Co 2p or the O 1s RC can be used for $I_{CoO_2}(\theta)$. See Section 6 of SI for more details on decomposition of valence band spectra.

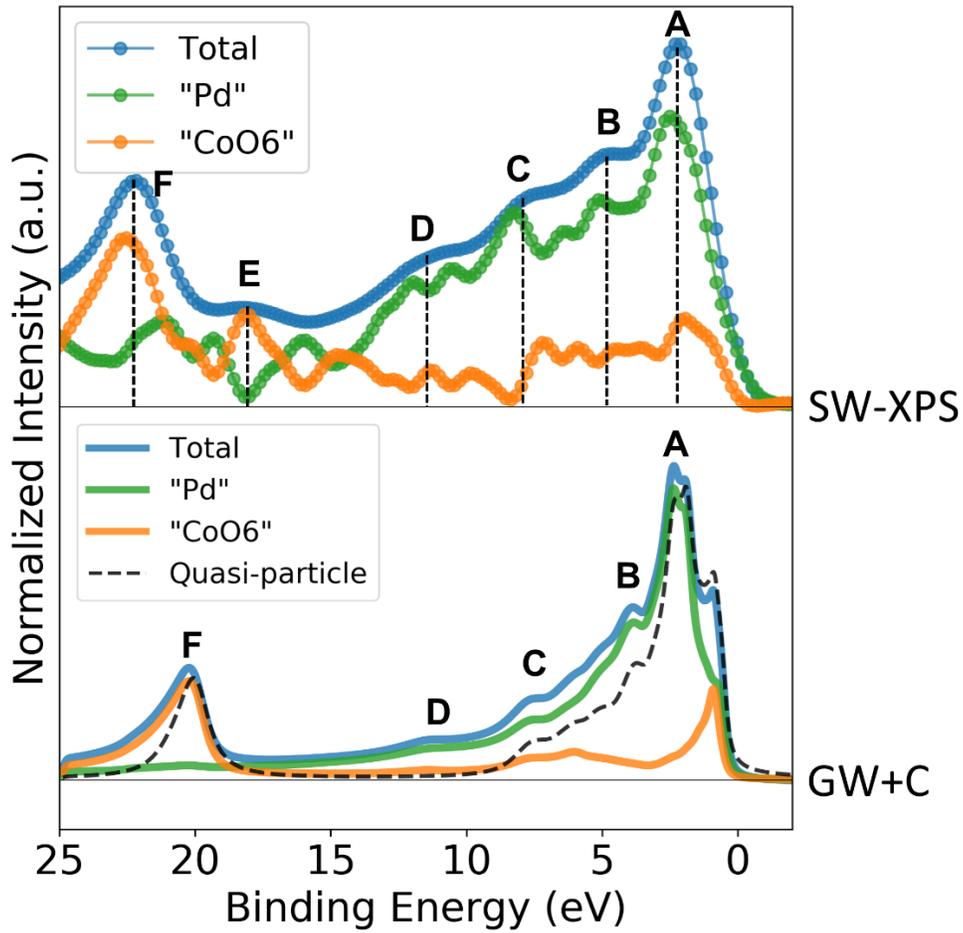

Figure 4: Layer-resolved valence band contributions of PdCoO$_2$ from standing wave X-ray photoemission spectroscopy (top panel) and GW+cumulant theory calculations (bottom panel). The dashed lines indicate the positions of the peaks in the experimental valence band spectrum.

The total and the layer-resolved valence band spectra are shown in the top panel of Figure 4. The total spectrum (blue curve) exhibits a large peak (labelled "A" in Figure 3) at a binding energy of ~2.5 eV. By looking at the layer-specific spectra, the dominant contribution to this peak derives from the Pd layer (green curve), but the CoO$_2$ spectrum (orange curve) also exhibits a smaller peak at the same binding energy. At higher binding energies, a series of three shoulder-like features is observed in the total spectrum (labelled "B", "C" and "D") which derive from the Pd layer. At a binding energy of ~18 eV, the total VB spectrum exhibits a small peak (labelled "E") originating from the CoO$_2$ layer. Finally, the large peak at ~23 eV (labelled "F") is also found to derive from the CoO$_2$ layer. To qualitatively confirm the layer-specific spectral features of the VB spectrum, we also performed resonant photoemission spectroscopy (RESPES) measurements, shown in Figure S3. In RESPES, the photon energy is tuned to a resonance condition, in this case the Pd L$_3$-edge (3173 eV), which results in additional photoemission intensity in VB spectrum because of final-state interference between the direct photoemission process and the autoionization process[35,36]. By subtracting the off-resonance spectrum collected at 3182 eV, it is thus possible to extract the contribution from the Pd 4d states. Figure S3 shows that the RESPES difference spectrum is in good qualitative agreement with the VB contribution from the Pd layer shown in Figure 4.

The assignment of the features in the experimental valence band spectrum is considered next. In the bottom panel of Figure 4, the theoretical layer-resolved valence band photoemission spectrum of PdCoO$_2$ from GW+C theory is shown. The quasiparticle part of the total spectrum (dashed line), which results from the direct photoemission process without the additional shake-up of an excitation, is also plotted. Peaks "A", "B", "C", and "F" of the experimental spectrum can be assigned to features in the quasiparticle spectrum, and, thus, result from band structure effects. By projecting the quasiparticle spectrum onto different atomic orbitals (see Figure S4), we find that the Pd layer spectrum is dominated by Pd 4d states with a slight admixture of Pd 5s orbitals. The peak in the CoO$_2$ spectrum near the Fermi level originates from antibonding combinations of Co 3d and O 2p orbitals. In contrast, the broad peak in the CoO$_2$ spectrum near 5 eV derives from Co 3d and O 2p states that form bonding states. Finally, the peak "F" is found to derive from O 2s states. Directly at the Fermi level, states in the Pd layer provide the dominant contribution to the spectrum, which is consistent with ARPES measurement reported in literature[5,17].

Based on the quasiparticle band structure, no photoemission intensity over a wide binding energy range from 8 eV to 18 eV is expected. Therefore, band structure theory is unable to account for the features "D" and "E" in the measured VB spectrum. However, the full GW+C VB spectrum does exhibit a peak that coincides with the binding energy of the feature "D" demonstrating that this peak is a satellite caused by many-electron interactions.

In order to determine the physical origin of this satellite peak, we have analyzed all the calculated valence band spectral functions on the finite k-point grid used in our calculations and identified those that contribute most strongly to the satellite intensity between 10 and 11 eV below the Fermi level, see Figure 5(a). Here, the satellite intensity is defined as the difference between the full spectral function and the quasiparticle part of the spectral function. In Figure 5(a), the states at each *k*-point are labeled using numerical indices in the order of increasing energy, with state 1 being the lowest energy state included in the calculation, *i.e.*, excluding the core electrons contained in the pseudopotentials. Spectral functions of degenerate states, identical to those shown in Figure 5(a), are omitted for clarity. Examination of the spectral functions shown in Figure 5(a) indicates that peak "D" in the angle-integrated valence band spectrum has contributions from different spectral functions. In particular, some of the contributing spectral functions, *e.g.* state 23 at $k = (0.0, 0.375, 0.0)$ (green curve), look similar to the Co 2p core level spectrum, whereas others, *e.g.* state 19 at (0.0, 0.0, 0.0) (blue curve), look similar to the Pd 3d core level. (The coordinates of the *k*-points are given as fractions of the reciprocal lattice vectors.)

Surprisingly, state 23 at $k = (0.0, 0.375, 0.0)$, whose spectral function looks similar to the Co 2p core level, is in fact predominantly localized on the Pd atoms based on a Löwdin analysis[37]. In order to understand this observation, we have studied the correlation between the atomic orbital character of a valence state and the satellite structure of its spectral function. In Figure 5(b), the intensity of the satellite feature located 10 eV below the quasiparticle peak is plotted against the Co 3d character of the state for all valence band states. It is then clear that states with greater Co 3d character are in general more strongly coupled to the 10 eV loss feature, which we have previously assigned to a charge transfer excitation in the $CoO_2$ layer. Similarly, it is possible to show that states with greater Co 3d character are also more strongly coupled to the loss feature at 3 eV, and that the loss feature at 8 eV, observed in the Pd 3d core level spectrum, is relatively weak in all cases.

Overall, the following picture emerges: the spectral functions of states localized in the $CoO_2$ layer are similar to the Co 2p core level spectrum, and the spectral functions of states localized in the Pd layer are similar to the Pd 3d core level spectrum. This finding is somewhat surprising as the creation of a hole induces a long-ranged Coulomb potential that is also experienced by electrons on neighboring layers. Spectral functions of states hybridized across both layers exhibit both types of loss features, but the coupling to the excitations in the $CoO_2$ layer is significantly stronger. As a consequence, states that are mostly (but not fully) localized on a Pd-layer still have a strong satellite peak arising from coupling to $CoO_2$ layer charge-transfer excitons.

These insights into the spectral functions of individual valence states allow us to understand the origin of peak "D" in the experimental valence band photoemission spectrum. In general, the valence states that contribute to peak "A" in the quasiparticle spectrum are hybridized across both the Pd and the $CoO_2$ layers. Due to the presence of Co 3d character, the spectral functions of these states exhibit a pronounced loss feature ~10 eV below the quasiparticle peak, or ~11 eV below the Fermi level corresponding to the binding energy of peak "D". However, in a photoemission experiment with ~3 keV photon energy, most of the spectral intensity arising from these hybridized states is derived from the Pd layer, because the atomic Pd 4d photoionization cross-section is significantly higher than the atomic Co 3d or O 2p photoionization cross-sections at this photon energy. We note that the use of atomic photoionization cross-sections even for delocalized valence states is well justified in X-ray photoemission, as discussed in Ref [38].

Finally, we note that the GW+C valence band spectrum does not exhibit a clear peak in the binding energy region corresponding to feature "E". Inspection of individual GW+C spectral functions of valence states does reveal satellite features at the same energy as feature "E", see Figure 5, but they are too weak and spread over too wide an energy window to give rise to a distinct peak in the angle integrated spectrum. The underestimation of the height of peak "E" by GW+C is likely a consequence of the neglect of extrinsic losses which transfer spectral weight from quasiparticle to satellite features.

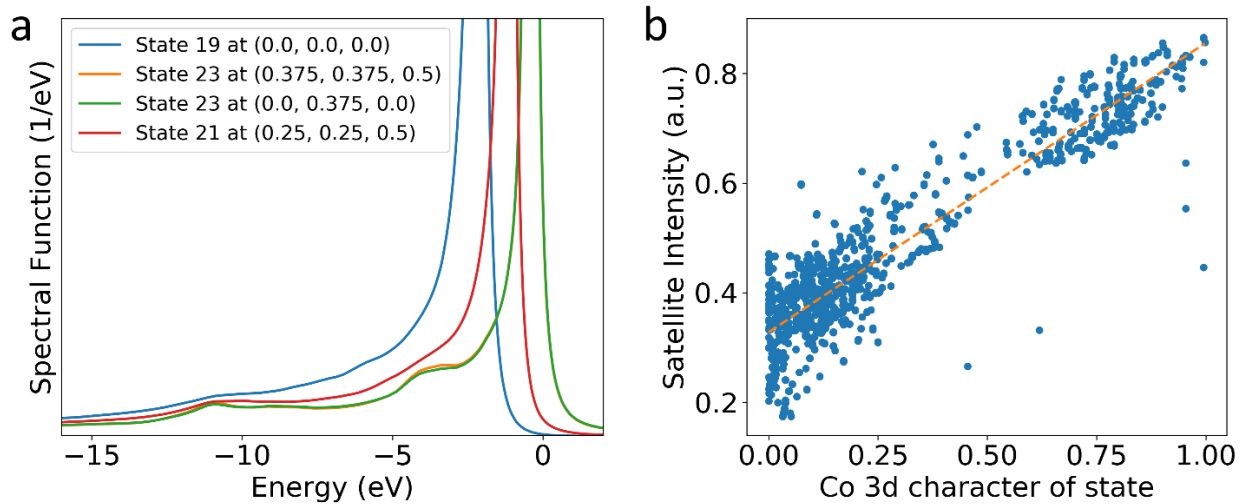

Figure 5: (a) Selected GW+C spectral functions that contribute strongly to the satellite intensity between 10 and 11 eV below the Fermi level in the theoretical photoelectron spectrum of PdCoO$_2$. (b) Analysis of valence band spectral functions in PdCoO$_2$. The scatter plot demonstrates the correlation between the intensity of the satellite 10 eV below the quasiparticle peak and the Co 3d character of the valence state. The correlation is also demonstrated by the linear fit (the orange dashed line).

In summary, we have used standing wave photoemission spectroscopy with tender/hard X-rays to obtain layer-specific valence band signatures. In conjunction with advanced many-body calculations, we provide a comprehensive understanding of the electronic structure of PdCoO$_2$. Using the GW+C calculations we are able to interpret not only quasiparticle features in this material, but also the nature of electron-plasmon and electron-exciton couplings. Moreover, the natural anisotropy of delafossite materials, which originates from the stacking of metallic Pd layers and Mott-insulating CoO$_2$ layers, also drastically affects the electron-electron as well as electron-plasmon interactions in this material. In particular, our study reveals that many-electron interactions in this material are highly layer sensitive: holes in the CoO$_2$ layer are strongly coupled to charge transfer excitations in the same layer, while holes in the Pd layer interact with plasmons in that layer. Holes in states that are hybridized across both layers couple to excitations in both layers, with the strength of the photoemission satellites being proportional to the projection of the state onto a given layer. However, in general, the charge transfer satellites are much more intense than the plasmon satellites. As a result, those states near the Fermi level that are mostly localized on the Pd layer, but also contain minor contributions from the CoO$_2$ layer, show clear satellites originating from coupling to charge-transfer excitons in the CoO$_2$ layer. Further investigations are required to establish whether this coupling has implications on the conduction of electrons in PdCoO$_2$ in the direction perpendicular to the layers, which has been an intriguing subject of recent studies[39,40]. As a whole, this work has expanded the current level of our understanding of the fascinating many-electron effects in layered quantum materials and highlighted the utility of the methodologies applied (standing wave photoemission, GW+cumulant calculations, cluster model calculations) for investigations of materials that host both weakly and strongly correlated electrons.

**Methods**

**Thin film deposition and structural characterizations.** $PdCoO_2$ thin film samples were grown by using molecular beam epitaxy (MBE) on $Al_2O_3$ sapphire (0001) substrates. Details about MBE growth can be found in our previous publication[41]. High-resolution X-ray diffraction and reflectivity measurements were performed by using a Panalytical X'Pert diffractometer with monochromated Cu $K_{\alpha1}$ X-ray beam ($\lambda$ = 1.5406 Å). To characterize the sample quality, we performed both X-ray reflectivity (XRR) and high-resolution X-ray diffraction (XRD) measurements using a monochromated Cu $K_{\alpha1}$ hard X-ray beam. Both XRR and XRD (Figure S1 in the Supporting Information) data show clear thickness Kiessig fringes, indicating the high crystal quality of the sample. The $PdCoO_2$ film is oriented along the (0001) direction with a corresponding lattice constant of 5.9 Å, i.e. the Pd and $CoO_2$ layers are parallel to the sample surface. The sample quality and the measured lattice parameter are comparable to previous literature reports for samples grown by either MBE[42] or pulsed laser deposition (PLD)[43].

**Tender/hard X-ray and standing wave photoemission spectroscopy.** Tender/hard X-ray and standing wave photoemission spectroscopy (SW-XPS) was performed at HAXPES endstation of beamline 9.3.1, Advanced Light Source (ALS), Lawrence Berkeley National Laboratory (LBL). Core level and valence band spectra were measured by using 4 keV photon energy at ~1° incident angle (so-called Henke peak[44]) to increase the signal intensity and improve signal-to-noise ratio. SW-XPS was performed at room temperature with photon energy tuned to 3173 eV exciting (0003) reflection at Pd $L_3$-resoance to enhance reflectivity. Core levels (Pd 3d, Pd 3p, Co 2p and O 1s) as well as valence band spectra were collected at each incident angles ranging from 18.5° to 21° with a step size of 0.02°. We also performed resonant photoemission spectroscopy of valence band spectra with photon energy ranging from 3168 eV to 3182 eV (Pd $L_3$-resoance). All measurements were performed with p-polarized X-rays with total energy resolution of ~600 meV. The sample was measured at room temperature in ultra-high vacuum (~$10^{-9}$ Torr).

**Cluster model calculations**. The core-level photoemission spectra were simulated using cluster model calculations. In this approach $PdCoO_2$ was treated as two separate basic units, a $PdO_2$ dumbbell and an edge-sharing $CoO_6$ octahedron, which whose spectra were determined in separate calculations. The Hamiltonian includes the six p- or ten d-orbitals of the core-level involved in the photoemission process and the ten transition metal valence d-orbitals, as well as the corresponding ligand valence orbitals. In the case of the PdO$_2$ dumbbell, these are the Pd 3d, Pd 4d, and O 2p orbitals, and for the CoO$_6$ octahedra, the Co 2p, Co 3d, and O 2p orbitals are included. In an ionic description, the Pd ion is monovalent (*i.e.*, Pd$^{1+}$: 4d$^9$) and the Co ion is trivalent (*i.e.*, Co$^{3+}$: 3d$^6$). The ground-state wave function of each system is expanded into a linear combination of the ionic configuration $|d^n\rangle$ plus the charge-transfer configurations such as $|d^{n+1}\underline{L}\rangle$, $|d^{n+2}\underline{L}^2\rangle$, etc., where $\underline{L}$ denotes a symmetry-adapted O 2p ligand hole.

The Hamiltonian also includes electron-electron interactions and final-state charge-transfer effects that are important to give a quantitative description of photoemission spectra. The relevant parameters are the Coulomb repulsion U, the attractive core-hole potential Q, the p-d charge-transfer energy Δ, and the charge-transfer integral T$_\sigma$. Here, Δ and U are reported with respect to the center of gravity of the multiplet of each configuration. Additional multiplet splitting is given in terms of the crystal field parameter 10Dq.

In order to include non-local contributions to the screening of the photoemission core-hole in the main CoO$_6$ unit, we also considered an additional set of ligand orbitals that describe the effect of the neighboring octahedra. In this case, Δ' and T' denote the charge-transfer energy and hybridization between the Co 3d orbitals and this new set of orbitals. Inclusion of these additional orbitals results in an additional set of charge-transfer configurations for the CoO$_6$ system: d$^6$, d$^7\underline{L}$, d$^7\underline{B}$, d$^8\underline{L}^2$, d$^8\underline{L}\underline{B}$, d$^8\underline{B}^2$, etc., where $\underline{L}$ denotes a hole in the local O 2p orbitals of the octahedron, and $\underline{B}$ denotes a hole in the remote ligand orbitals.

For the CoO$_6$ calculation, the values of the parameters are U = 6.5 eV, Q = 7.8 eV, Δ = 1 eV, and T$_\sigma$ = 3.0 eV. The crystal field 10Dq was set to 2.5 eV, which is close to the value calculated in DFT. For the effective bath, the parameters are Δ' = -2.0 eV and T' = 1.0 eV. The calculated ground state of the system is low-spin (<S> = 0), and is composed of 36% 3d$^7\underline{L}$, 35% 3d$^6$, 16% 3d$^7\underline{B}$, 6.5% 3d$^8\underline{L}^2$, 5.1% 3d$^8\underline{L}\underline{B}$, as well as smaller contributions from higher occupied charge-transfer states. This leads to a mean occupation of 6.79 electrons in the Co 3d orbitals. The relatively high configuration weight of the charge-transfer configurations compared to the ionic state 3d$^6$ shows the strong degree of covalence in the CoO$_6$ octahedra, as well as the importance of the additional effective bath.

After the Hamiltonian is diagonalized and the ground-state is obtained, the corresponding spectral weight is obtained for the core level photoemission. The calculations were performed using the Green function formalism as implemented in Quanty[45].

**GW+cumulant calculations.** To model the valence band electronic structure of $PdCoO_2$, first-principles GW+C calculations were carried out. First, DFT calculations were performed using the Quantum Espresso software package[46]. The Perdew–Burke–Ernzerhof (PBE) exchange-correlation functional, scalar-relativistic optimized norm-conservinng Vanderbilt pseudopotentials, a 60.0 Rydberg wave-function cutoff, and an 8x8x4 k-point grid were used[47,48]. Next, based on the PBE starting point, full frequency G0W0 calculations were carried out as implemented in the BerkeleyGW software package[49]. The frequency-dependent dielectric matrix was calculated within the random phase approximation. Frequencies from 0 eV to 60 eV were sampled on a uniform grid with a spacing of 0.08 eV, and frequencies between 60 eV and 400 eV were sampled with a non-uniform grid, where each successive frequency step is increased by 0.15 eV. A total of 400 bands were included in the calculation. We then computed the electronic self-energy on a frequency grid that ranges from 22.1 eV below the mean-field Fermi level to 5.15 eV above the mean-field Fermi level and contains 110 equally spaced frequency points. The static remainder correction in the Coulomb hole term was included, as described in Ref. [50]. The GW+C spectral functions at each k-point are constructed from the G0W0 self-energies following the method described in Ref. [28]. Finally, the Brillouin-zone integrated photoelectron spectrum was calculated as a cross-section weighted sum of the individual GW+C spectral functions, where the cross-section for each state at each k-point was obtained by combining the projections of that state onto local atomic orbitals (obtained from a Löwdin analysis) with tabulated atomic orbital cross-sections from Ref. [51].

We also used the GW+C approach to calculate approximate Co 2p and Pd 3d core-level spectral functions. Of course, a direct calculation of these spectral functions is not possible as we are using a pseudopotential approach. However, it is possible to identify specific valence states which are strongly localized on either the Co or the Pd atoms. It is reasonable to expect that the satellite structure of these valence states is very similar to that of the corresponding core levels. To construct core-level spectral functions from the computed GW+C valence electron spectral functions, two copies of the same valence spectral functions were added together with one copy being shifted by the atomic spin-orbit splitting and weighted by the theoretical intensity ratio from the multiplicities of the j-states.

**ACKNOWLEDGEMENTS**

This research used HAXPES endstation at beamline 9.3.1 of the Advanced Light Source, a U.S. DOE Office of Science User Facility under contract no. DE-AC02-05CH11231. Q.L. was supported by ALS collaborative fellowship program. Q.L. and W.C.C. gratefully acknowledge financial support through the Department of Energy, Office of Basic Energy Sciences, Division of Materials Sciences and Engineering under contract no. DE-AC02-76SF00515. J.L. and J.M.K. acknowledge funding from EPSRC under Grant No. EP/R002010/1 and from a Royal Society University Research Fellowship (URF\R\191004). This work used the ARCHER UK National Supercomputing Service via J.L.'s membership of the HEC Materials Chemistry Consortium of UK, which is funded by EPSRC (EP/L000202). G.R. and S.O. are funded by National Science Foundation (NSF) Grant No. DMR2004125 and Army Research Office (ARO) Grant No. W911NF-20-1-0108. M.B. acknowledges the U.S. Department of Energy, Office of Science, Basic Energy Sciences, Materials Sciences and Engineering Division. H.P.M. has been supported for salary by the U.S. Department of Energy (DOE) under Contract No. DE-SC0014697.